\documentclass[journal]{IEEEtran}

\ifCLASSINFOpdf
  \usepackage[pdftex]{graphicx}
  \DeclareGraphicsExtensions{.pdf,.jpeg,.png}
\else
  \usepackage[dvips]{graphicx}
\fi
\usepackage{epstopdf}
\usepackage[cmex10]{amsmath}
\usepackage{amssymb}
\usepackage{wasysym}
\usepackage{algorithm}
\usepackage{algorithmic}
\usepackage{array}
\usepackage{cite}
\usepackage{color}
\usepackage{url}

\usepackage{epsfig,latexsym}
\usepackage{flushend}
\usepackage{cite}
\usepackage{verbatim}
\usepackage{amsopn}
\usepackage{booktabs}

\usepackage{stfloats}
\usepackage{hyperref}
\usepackage{graphicx}
\usepackage{subfigure}

\begin{document}

\title{Sparse Channel Estimation and Hybrid Precoding Using Deep Learning for Millimeter Wave Massive MIMO}

\author{Wenyan~Ma,~\IEEEmembership{Student~Member,~IEEE}, Chenhao~Qi,~\IEEEmembership{Senior~Member,~IEEE}, \\ Zaichen Zhang,~\IEEEmembership{Senior~Member,~IEEE}, and Julian Cheng,~\IEEEmembership{Senior~Member,~IEEE}
\thanks{This work was supported in part by National Natural Science Foundation of China under Grant 61871119 and 61960206005, by Natural Science Foundation of Jiangsu Province under Grant BK20161428, by National Key Research and Development Plan Project under Grant 2018YFB1801101, and by the Fundamental Research Funds for the Central Universities. (\textit{Corresponding author: Chenhao~Qi})}
\thanks{Wenyan~Ma, Chenhao~Qi and Zaichen Zhang are with the School of Information Science and Engineering, Southeast University, Nanjing 210096, China (Email: \{qch,zczhang\}@seu.edu.cn).}
\thanks{Julian Cheng is with School of Engineering, The University of British Columbia, Kelowna, BC, Canada (Email: julian.cheng@ubc.ca).}
}

\markboth{Submitted to IEEE Transactions on Communications}
{Shell \MakeLowercase{\textit{et al.}}: Bare Demo of IEEEtran.cls for Journals}

\maketitle

\begin{abstract}
Channel estimation and hybrid precoding are considered for multi-user millimeter wave massive multi-input multi-output system. A deep learning compressed sensing (DLCS) channel estimation scheme is proposed. The channel estimation neural network for the DLCS scheme is trained offline using simulated environments to predict the beamspace channel amplitude. Then the channel is reconstructed based on the obtained indices of dominant beamspace channel entries. A deep learning quantized phase (DLQP) hybrid precoder design method is developed after channel estimation. The training hybrid precoding neural network for the DLQP method is obtained offline considering the approximate phase quantization. Then the deployment hybrid precoding neural network (DHPNN) is obtained by replacing the approximate phase quantization with ideal phase quantization and the output of the DHPNN is the analog precoding vector. Finally, the analog precoding matrix is obtained by stacking the analog precoding vectors and the digital precoding matrix is calculated by zero-forcing. Simulation results demonstrate that the DLCS channel estimation scheme outperforms the existing schemes in terms of the normalized mean-squared error and the spectral efficiency, while the DLQP hybrid precoder design method has better spectral efficiency performance than other methods with low phase shifter resolution.

\end{abstract}

\section{Introduction}
Due to the rich bandwidth resources of the millimeter wave (mmWave), mmWave communication has attracted broad attention and become an important technology in future wireless communication systems \cite{heath2016overview,li20155G}. When operating at high frequency, the mmWave signal experiences high path loss. Fortunately, this challenge can be overcome by directional beamforming with a massive  multi-input multi-output (MIMO) antenna array. Since mmWave bands have short wavelengths, large antenna arrays can be packed into small form factors~\cite{amadori2015low}.

Due to the large antenna arrays of mmWave communications, channel estimation requires a large number of time slots as overhead. Note that the mmWave channels have sparsity feature in the beamspace domain with hybrid precoding~\cite{dai2016estimation}. Although the beamspace is typically addressed in the mmWave lens antenna arrays, we can also obtain the beamspace channel with hybrid precoding  by introducing a dictionary matrix consisting of column steering vectors. Several channel estimation schemes have been proposed to explore the beamspace channel sparsity. For examples, a distributed grid matching pursuit (DGMP)  channel estimation scheme was proposed~\cite{dai2016estimation}, where the dominant entries of the line-of-sight (LOS) channel path were detected and updated iteratively; an orthogonal matching pursuit (OMP)  channel estimation scheme was proposed to detect the dominant entries of multiple channel paths~\cite{venugopal2017time}; a simultaneous weighted orthogonal matching pursuit (SWOMP)  channel estimation scheme was proposed~\cite{javier2018frequency}, where the frequency-selective mmWave channels were considered based on the OMP method. However, these compressed sensing (CS) channel estimation schemes estimate the dominant beamspace channel entries sequentially and greedily, which cannot guarantee the global optimality~\cite{ma2018beamspace}.

After the channel estimation of mmWave communications, hybrid precoding consisting of analog precoding and digital precoding is usually adopted. Analog precoding aims to form directional beams using phase shifter network, while digital precoding is designed to mitigate interference of multiple data streams. Several hybrid precoding methods have been proposed for single-user multi-stream mmWave communication systems. For examples, a hybrid precoding algorithm was proposed~\cite{ayach2014spatially}, where the analog precoding problem was formulated as a sparse reconstruction problem and the OMP method was adopted; to avoid the greed of the OMP method, the alternating minimization method was used~\cite{yu2016alternating}, where the hybrid precoding problem was designed as a matrix decomposition problem and the  analog precoder and digital precoder were optimized alternately; to reduce the computational complexity of the alternating minimization method, the hierarchical codebook was used to obtain multiple beams and then form the analog precoding~\cite{xiao2016hi,chen2020two}.

In the multi-user multi-stream mmWave communication systems, the base station (BS) transmits multiple data steams to serve all users simultaneously. To improve the spectral efficiency, the beamsteering codebook based on steering vectors was used to formulate the analog precoder vectors and the digital precoder was designed~\cite{alk2015limited}. To consider the hardware constraint of the limited phase shifter resolution, beam allocation for multiple users was considered~\cite{sun2019beam}, where the discrete fourier transformation (DFT) codebook was adopted for analog precoding and the phase shifter resolution must be proportional to the number of antennas. To remove the constraint that the phase shifter resolution was related to the number of antennas, a quantized angle linear search (QALS) precoding scheme was proposed~\cite{zhao2017multi}, where  the angular domain was quantized according to the limited resolution of phase shifters and a linear search method was used to obtain the optimal analog beamforming vectors aligning with the dominant channel paths.  However, these hybrid precoding schemes design the analog precoder using the steering vectors of quantized angles, which is heavily constrained by the resolution of phase shifters. When the mmWave system is equipped with low resolution phase shifters, there is a small number of available steering vectors of quantized angles. Since the angles of arrival (AoAs) of channel paths are randomly distributed, it cannot guarantee that the precoding based on these limited steering vectors can always have the high beamforming gain. Therefore these hybrid precoding schemes may have unsatisfactory spectral efficiency performance if none of these limited steering vectors can be aligned with the AoAs well.

Recently, the application of deep learning to mmWave communications has received much attention owing to the capability of deep learning to solve complicated  nonlinear problems~\cite{ye2018power,qin2019deep,liu2019user}. For examples, a machine learning based beam prediction scheme was proposed~\cite{wang2018mmwave}, where the machine learning tools and situational awareness were combined to learn the beam information (power, optimal beam index, etc) from past observations; a learned denoising based approximate message passing network was proposed to estimate the mmWave communication system with lens antenna array~\cite{he2018deep}, where the noise term was detected and removed to estimate the channel. However, channel estimation for mmWave massive MIMO systems with hybrid precoding was not considered~\cite{he2018deep}. Besides, a deep learning based beamforming design method was proposed~\cite{lin2019beamforming}, where a beamforming neural network was trained to learn how to optimize the beamformer for maximizing the spectral efficiency;  a deep
reinforcement learning hybrid precoding method was proposed~\cite{wang2019precoder}. However, both these two deep learning hybrid precoder design methods neglect the constraint of limit resolution of phase shifters.

In this paper, we investigate sparse channel estimation and hybrid precoding considering the limited resolution of phase shifters for multi-user mmWave massive MIMO systems. The paper has the following two main contributions.

1) We propose a deep learning compressed sensing (DLCS) channel estimation scheme for the multi-user mmWave massive MIMO systems. The DLCS scheme consists of  beamspace channel amplitude estimation and channel reconstruction.  In the offline training stage, we train the channel estimation neural network (CENN)  using the simulated environment based on the mmWave channel model. Then  in the online deployment stage, the correlation between the received signal vectors and the measurement matrix is fed into the trained CENN to predict the beamspace channel amplitude. Afterwards, the indices of dominant entries of beamspace channel are obtained, based on which the channel can be reconstructed. Unlike the existing work that estimates the dominant beamspace channel entries sequentially~\cite{dai2016estimation,venugopal2017time,javier2018frequency}, we estimate dominant entries simultaneously, which will be shown to have better channel estimation performance.

2) We propose a deep learning quantized phase (DLQP) hybrid precoding method for the multi-user mmWave massive MIMO systems. In the DLQP method, we first design the analog precoder and then the digital precoder.  In the offline training stage, we obtain the training hybrid precoding neural network (THPNN)  using the estimated channel vector and real channel vector of each user, where the approximate phase quantization is considered. Then  in the online deployment stage, we obtain the deployment hybrid precoding neural network (DHPNN) by replacing the approximate phase quantization in the THPNN with ideal phase quantization, where the estimated channel vector of each user is fed into the DHPNN to obtain the analog precoding vector. Afterwards, the analog precoding matrix is obtained by stacking the analog precoding vectors of all users, based on which the digital precoding matrix can be calculated by zero-forcing (ZF).

The rest of the paper is organized as follows. In Section II, we introduce the system model and formulate the problem of channel estimation for the multi-user mmWave massive MIMO systems with hybrid precoding. In Sections III, we propose the DLCS channel estimation scheme. In Section IV, we develop the DLQP hybrid precoder design method. The simulation results are provided in Section V. Finally, Section VI concludes the paper.

We use the following notations. Symbols for  vectors (lower case) and matrices (upper case) are in boldface.  $(\cdot)^T $, $(\cdot)^* $,  $(\cdot)^H $, and $(\cdot)^{-1} $ denote the transpose, conjugate,  conjugate transpose (Hermitian), and inverse, respectively. We use $\boldsymbol{I}_{K}$ to represent identity matrix of order $K$. The set of $P\times{Q}$ complex-valued matrices and real-valued matrices are denoted by $\mathbb{C}^{P\times{Q}}$ and $\mathbb{R}^{P\times{Q}}$, respectively. We use $\mathbb{E}\{\cdot\}$ to represent expectation.  The $l_2$-norm of a vector and Frobenius norm of a matrix are denoted by $\|\cdot\|_2$ and $\|\cdot\|_F$, respectively. We use $\boldsymbol{a}[p]$ to denote the $p$th entry of $\boldsymbol{a}$. Complex Gaussian distribution is denoted by $\mathcal{CN}$. We use $| \cdot|$ to denote the absolute value. $\rm {Im}(\boldsymbol{a})$ and $\rm{Re}(\boldsymbol{a})$ denote the imaginary and real parts of $\boldsymbol{a}$, respectively.

\section{System Model and Problem Formulation}
We first introduce the system model of multi-user mmWave massive MIMO. Then  the channel estimation problem is formulated as a CS problem to estimate the sparse channel in the beamspace.

\subsection{System Model}
\begin{figure}[!t]
\centering
\includegraphics[width=90mm]{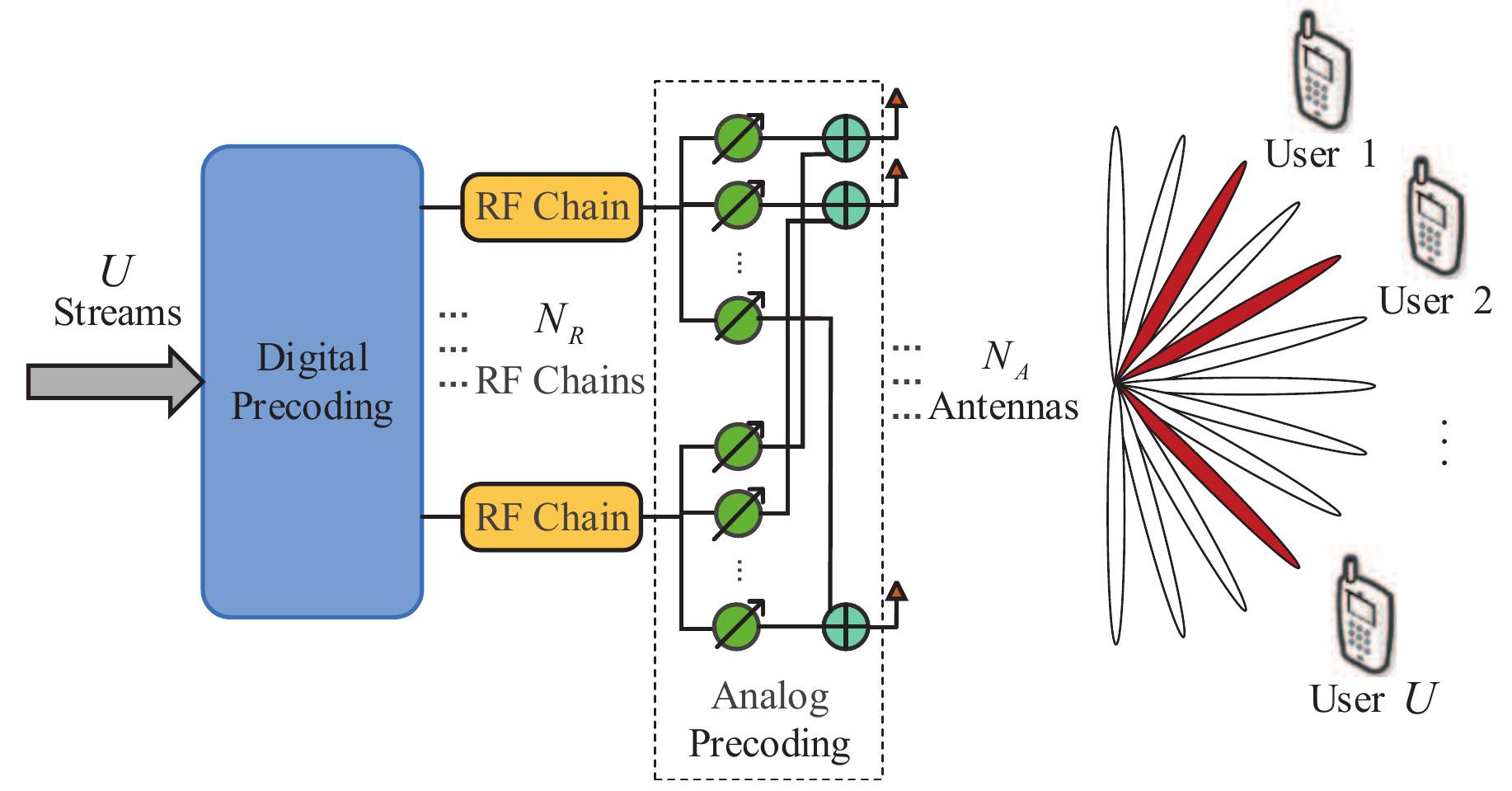}
\caption{Block diagram of downlink transmission in the multi-user mmWave massive MIMO system}
\label{FIG1}
\end{figure}

We consider a downlink multi-user mmWave massive MIMO communication system that comprises a BS and $U$  users with single antenna, as shown in Fig.~\ref{FIG1}. The BS is equipped with a uniform linear array (ULA)~\cite{heath2016overview}. Note that the present method can be generalized to other array structures. Let $N_A$ and $N_R$ denote the numbers of antennas and RF chains at the BS, respectively. Hybrid precoding is typically adopted, where the number of antennas is much larger than that of RF chains, i.e., $  N_A \gg N_R$~\cite{li20155G}. We consider the orthogonal multiple access, where the number of active users simultaneously connected with the BS is no larger than the number of RF chains, i.e., $U \leq N_R$~\cite{xiao2016hi}. If $U < N_R$, the BS will only turn on $U$  RF chains to  serve the $U$ users simultaneously and turn off $N_R-U$ RF chains, which will save the power consumed at the BS.

For downlink transmission, the BS performs hybrid precoding, which consists of baseband digital precoding and RF analog precoding~\cite{sun2019beam}. The received signal of all $U$ users, denoted by $\boldsymbol{y}^{\rm{dl}}\in{\mathbb{C}^{U}}$, can be represented as
\begin{equation}\label{UpSig}
\boldsymbol{y}^{\rm{dl}}=\boldsymbol{H} \boldsymbol{F}_R \boldsymbol{F}_B \boldsymbol{s} + \boldsymbol{n}
\end{equation}
where $\boldsymbol{F}_R\in{\mathbb{C}^{N_A\times{U}}}$ and $\boldsymbol{F}_B\in{\mathbb{C}^{U\times{U}}}$  denote the analog precoder and digital precoder, respectively. To normalize the power of the hybrid precoder, we set $\| \boldsymbol{F}_R \boldsymbol{F}_B \| _F ^2=U$. We denote  the signal vector by $\boldsymbol{s}\in{\mathbb{C}^{U}}$ satisfying $\operatorname{E}\{\boldsymbol{ss}^{H}\}=\boldsymbol{I}_{U}$ and  additive white Gaussian noise (AWGN) vector by $\boldsymbol{n} \in{\mathbb{C}^{U}} $ satisfying $\boldsymbol{n} \sim \mathcal{CN}(0,\sigma^{2}\boldsymbol{I}_{U})$. The channel matrix for the BS and all users is denoted by
\begin{equation}
\boldsymbol{H} \triangleq \left[\boldsymbol{h}_{1},...,\boldsymbol{h}_{U}\right]^{T} \in{\mathbb{C}^{U\times{N_A}}}.
\end{equation}
There are different kinds of channel model in mmWave systems, such as the clustered mmWave channel model and the Saleh-Valenzuela mmWave channel model~\cite{li20155G,tao2019regularlized}. We choose the Saleh-Valenzuela mmWave channel model in our paper. The channel vector $\boldsymbol{h}_{u}\in{\mathbb{C}^{N_A}}$ for the BS and the $u$th user is represented as
\begin{equation}\label{hu}
\boldsymbol{h}_{u}=\sqrt{\frac{N_A}{L_u}}\sum_{i=1}^{L_u} \boldsymbol{h}_{u,i}=\sqrt{\frac{N_A}{L_u}}\sum_{i=1}^{L_u} g_{u,i} \boldsymbol{\alpha} (N_A,\theta_{u,i})
\end{equation}
where the channel vector, number of multiple channel paths,  and complex gain of the $i$th path are denoted by $\boldsymbol{h}_{u,i}$, $L_{u}$,  and $g_{u,i}$, respectively. Typically $\boldsymbol{h}_{u}$ consists of one  LOS path (the 1st channel path), and $L_u-1$ non-line-of-sight (NLOS) paths (the $i$th channel path for $2\leq i \leq L_u)$.  The steering vector $\boldsymbol{\alpha}(N,\theta)$ can be expressed as
\begin{equation}\label{alpha1}
\boldsymbol{\alpha}(N,\theta)=\frac{1}{\sqrt{N}}\left[1,e^{j\pi\theta},...,e^{j\pi\theta(N-1)}\right]^{T}.
\end{equation}
Denote the AoA for the $i$th path of the $u$th user by $\vartheta_{u,i}$, which is uniformly distributed over $[-\pi,\pi)$ \cite{dai2016estimation,chen2019beam}. Then we have $\theta_{u,i} \triangleq \sin{\vartheta_{u,i}}$ if the distance between adjacent two antennas at the BS is half-wave length~\cite{dai2016estimation}.

\subsection{Problem Formulation}
To design $\boldsymbol{F}_B$ and $\boldsymbol{F}_R$ for downlink data transmission, $\boldsymbol{H}$ should be estimated. Based on channel reciprocity, the estimate of downlink channel can be obtained by employing uplink channel estimation to estimate $\boldsymbol{H}$. Note that the proposed DLCS  channel estimation scheme can also be used for the downlink channel estimation. Since the BS usually has more computing power than each user in practice, we consider the uplink channel estimation where the neural network (NN) is trained and utilized for prediction  at the BS.  For uplink channel estimation, mutually orthogonal pilot sequences are transmitted by all users to distinguish different signals from different users for $K$ times. Denote the pilot matrix consisted of the $U$ mutually orthogonal pilot sequences from $U$ users by $\boldsymbol{P}\in{\mathbb{C}^{U\times{U}}}$. For the uplink pilot transmission, we use $K$ different  analog precoding matrices and digital precoding matrices, denoted by  $\boldsymbol{F}_R^k\in{\mathbb{C}^{N_A\times{N_R}}}$ and $\boldsymbol{F}_B^k \in{\mathbb{C}^{N_R\times{N_R}}} $, respectively, for $k=1,2,\ldots,K$. The pilot sequences received at the BS for the $k$th sending are given by
\begin{equation}\label{YU}
\boldsymbol{Y}_k^{\rm{ul}}=(\boldsymbol{F}_R^k \boldsymbol{F}_B^k)^T \boldsymbol{H}^T \boldsymbol{P} + (\boldsymbol{F}_R^k \boldsymbol{F}_B^k)^T \boldsymbol{N}_{k}
\end{equation}
where the AWGN matrix for the $k$th transmission is denoted by $\boldsymbol{N}_{k}$. Each entry of $\boldsymbol{N}_{k}$ obeys $\mathcal{CN}(0,\sigma^{2})$. Based on the orthogonality of $U$ mutually orthogonal pilot sequences, i.e., $\boldsymbol{P}\boldsymbol{P}^{H}=\boldsymbol{I}_{U}$, we multiply $\boldsymbol{Y}_k^{\rm{ul}}$ by $\boldsymbol{P}^{H}$ and obtain
\begin{equation}
\boldsymbol{R}_{k} \triangleq \boldsymbol{Y}_k^{\rm{ul}} \boldsymbol{P}^{H}=(\boldsymbol{F}^k)^T \boldsymbol{H}^T + \widetilde{\boldsymbol{N}}_{k}
\end{equation}
where
\begin{align}
\boldsymbol{F}^k &\triangleq \boldsymbol{F}_R^k \boldsymbol{F}_B^k \in{\mathbb{C}^{N_A\times{N_R}}}, \notag \\
\widetilde{\boldsymbol{N}}_{k} &\triangleq (\boldsymbol{F}_R^k \boldsymbol{F}_B^k)^T \boldsymbol{N}_{k}\boldsymbol{P}^{H} \in{\mathbb{C}^{N_R\times{U}}}.
\end{align}
After each user repeatedly transmits orthogonal pilot sequences $K$ times,  $\boldsymbol{R}_{k}$ for $k=1,2,\ldots,K$ can be stacked as
\begin{equation}
\boldsymbol{R}=[\boldsymbol{R}_{1}^T,\ldots,\boldsymbol{R}_{K}^T]^T = \boldsymbol{F}^T \boldsymbol{H}^T +\widetilde{\boldsymbol{N}}
\end{equation}
where
\begin{align}
\boldsymbol{F} &\triangleq [\boldsymbol{F}^1,\ldots,\boldsymbol{F}^K] \in{\mathbb{C}^{N_A\times{N_R K}}}, \notag \\
\widetilde{\boldsymbol{N}} &\triangleq [\widetilde{\boldsymbol{N}}_1^T,\ldots, \widetilde{\boldsymbol{N}}_K^T]^T \in{\mathbb{C}^{N_R K\times{U}}}.
\end{align}
Note that $N_A>N_R K$ since $  N_A \gg N_R$ and we need a small number of time slots for channel training. Denote the $u$th column of $\boldsymbol{R}$ by $\boldsymbol{r}_{u}$ for $u=1,2,\ldots,U$. Then $\boldsymbol{r}_{u}$ can be represented as
\begin{equation}\label{ru}
\boldsymbol{r}_{u}=\boldsymbol{F}^{T}\boldsymbol{h}_{u}+\widetilde{\boldsymbol{n}}_{u}
\end{equation}
where $\widetilde{\boldsymbol{n}}_{u}$ is the $u$th column of $\widetilde{\boldsymbol{N}}$.

\begin{figure*}[!t]
\centering
\includegraphics[width=130mm]{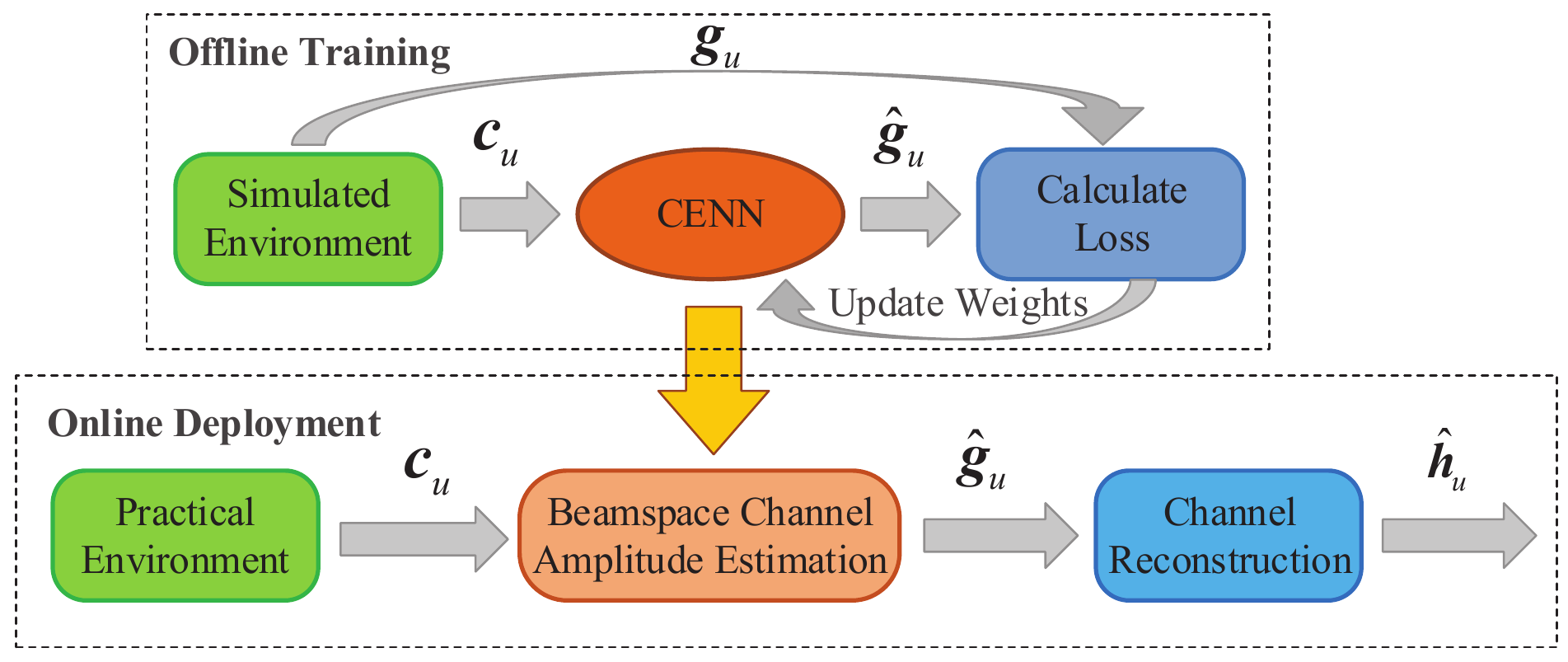}
\caption{Block diagram of the DLCS  channel estimation scheme: offline training and online deployment}
\label{FIG2}
\end{figure*}

Note that the mmWave channels have sparsity feature in the beamspace domain~\cite{dai2016estimation,javier2018frequency}. We define
\begin{equation}\label{ru2}
\boldsymbol{h}_{u}^b=\boldsymbol{A} \boldsymbol{h}_{u}
\end{equation}
as a beamspace channel vector where $\boldsymbol{A} \in{\mathbb{C}^{G\times{N_A}}}$ is the dictionary matrix consisted of $G$ column vectors $\boldsymbol{\alpha}(N_A,\phi_t)$, with $\phi_t \triangleq -1+2(t-1)/G$ representing the $t$th point of the angle grid. Note that the range of AoAs is quantified into $G$ grids for $t=1,2,\ldots,G$. Based on the fact that $\boldsymbol{A}^H\boldsymbol{A}=G\boldsymbol{I}_{N_A} / N_A$, eq. (\ref{ru2}) can be further rewritten as
\begin{equation}\label{ru3}
\boldsymbol{r}_{u}=\frac{N_A}{G}\boldsymbol{F}^{T} \boldsymbol{A}^H \boldsymbol{h}_{u}^b + \widetilde{\boldsymbol{n}}_{u}.
\end{equation}
Due to the sparse property of $\boldsymbol{h}_{u}^{b}$, eq. (\ref{ru3}) is essentially a sparse recovery problem, which can be tackled by CS techniques~\cite{qin2018sparse}. Note that the sparsity of $\boldsymbol{h}_{u}^{b}$ can be impaired by channel power leakage caused by the limited beamspace resolution of $\boldsymbol{A}$~\cite{brady2013beam}, which indicates that $\boldsymbol{h}_{u}^{b}$ is not perfectly sparse and  many  entries of $\boldsymbol{h}_{u}^{b}$ are small but nonzero. Sparse channel estimation schemes such as OMP and DGMP estimate the dominant beamspace channel entries in a sequential and greedy manner. However, they cannot guarantee the global optimality. Therefore, in the following we will propose a DLCS channel estimation scheme to estimate dominant beamspace channel entries simultaneously.

\section{DLCS Channel Estimation}
The proposed DLCS channel estimation scheme consists of beamspace channel amplitude estimation and channel reconstruction. The main idea of the DLCS scheme is  to  estimate first the beamspace channel amplitude using an offline-trained CENN, and then sort the estimated beamspace channel amplitude in descending order to select the indices of dominant entries, and finally reconstruct the channel according to the selected indices. The block diagram of the DLCS  scheme is illustrated in Fig.~\ref{FIG2}. The detailed steps of  the DLCS scheme are summarized in Algorithm~\ref{alg1}.

\subsection{Beamspace Channel Amplitude Estimation}
We define
\begin{equation}
  \boldsymbol{\Phi} \triangleq \frac{N_A}{G}\boldsymbol{F}^{T} \boldsymbol{A}^H \in{\mathbb{C}^{N_R K\times{G}}}
\end{equation}
as the measurement matrix in (\ref{ru3}). As shown in Algorithm~\ref{alg1}, we feed $\boldsymbol{\Phi}$ and $\boldsymbol{r}_{u}$ to obtain the estimate of $\boldsymbol{h}_{u}$, denoted by $\hat{\boldsymbol{h}}_{u}$, for $ u=1,2,\ldots,U$. The correlation vector between $\boldsymbol{\Phi}$ and $\boldsymbol{r}_{u}$, denoted by $\boldsymbol{c}_{u} \in{\mathbb{C}^{G}}$, can be expressed as
\begin{equation}\label{c}
\boldsymbol{c}_{u}=\boldsymbol{\Phi}^H \boldsymbol{r}_{u}.
\end{equation}

The sparse channel estimation schemes sequentially select the atoms, i.e., column vectors of $\boldsymbol{\Phi}$, which yield the greatest  correlation with $\boldsymbol{r}_{u}$. However, such greedy algorithms cannot guarantee the global optimality, which motivates us to use the NN to estimate the atoms simultaneously instead of sequentially.

As shown in Fig.~\ref{FIG2}, the beamspace channel amplitude estimation has two stages: the offline training of the CENN and its online deployment. The CENN is first trained offline and then used as the kernel of the beamspace channel amplitude estimation. The input of the CENN is $\boldsymbol{c}_{u}$. The amplitude of $\boldsymbol{h}_{u}^{b}$ can be denoted by
\begin{equation}\label{gu}
\boldsymbol{g}_{u}\triangleq \left[\left|\boldsymbol{h}_{u}^{b}[1]\right|,\left|\boldsymbol{h}_{u}^{b}[2]\right|,\ldots,\left|\boldsymbol{h}_{u}^{b}[G]\right|\right]^T \in{\mathbb{R}^{G}}.
\end{equation}
The output of the CENN is denoted by $\hat{\boldsymbol{g}}_{u}$ and is expected to be $\boldsymbol{g}_{u}$.

\begin{figure}[!t]
\centering
\includegraphics[width=90mm]{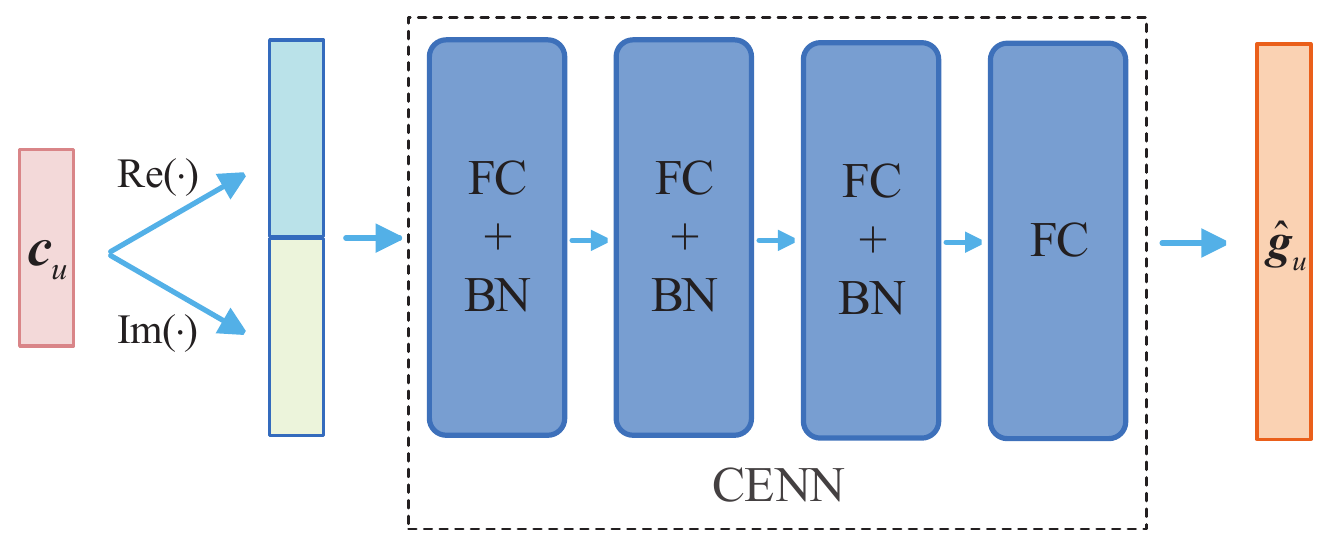}
\caption{Illustration of the CENN}
\label{FIG3}
\end{figure}

As illustrated in Fig.~\ref{FIG3}, the adopted CENN in this work consists of three hidden layers and a fully connected (FC) layer. Since the NN can only deal with the real number, the input of the CENN is a real-valued vector having $2G$ entries composed by the imaginary and real  parts of $\boldsymbol{c}_{u}$.  Each hidden layer includes an FC layer and a batch normalization (BN) layer. The numbers of neurons in these three hidden layers are set as 1,024, 512, and 256. The activation function adopted in the FC layer is the ReLU function, which can be represented as $f_{\rm Re}(x)=\max(0,x)$.

\begin{algorithm}[!t]
	\caption{ DLCS  Channel Estimation}
	\label{alg1}
	\begin{algorithmic}[1]
		\STATE \emph{Input:} $\boldsymbol{\Phi}$, $\boldsymbol{r}_{u}$, $J$.
        \STATE Initialization:  $\hat{\boldsymbol{h}}_{u}^{b}\leftarrow\boldsymbol{0}^G$.
        \vspace{5 pt}
        \STATE \emph{(Beamspace Channel Amplitude Estimation)}
        \STATE Obtain $\boldsymbol{c}_{u}$ via (\ref{c}).
        \STATE Input $\boldsymbol{c}_{u}$ to the  offline-trained CENN to get $\hat{\boldsymbol{g}}_{u}$.
        \vspace{5 pt}
        \STATE \emph{(Channel Reconstruction)}
        \STATE Obtain $\boldsymbol{\Gamma}$ based on $J$  dominant entries of $\hat{\boldsymbol{g}}_{u}$.
        \STATE Compute $\hat{\boldsymbol{h}}_{u}^{b}[\boldsymbol{\Gamma}]$ via (\ref{14}).
        \STATE Obtain $\hat{\boldsymbol{h}}_{u}$ according to (\ref{15}).
        \vspace{5 pt}
        \STATE \emph{Output:} $\hat{\boldsymbol{h}}_{u}$.
	\end{algorithmic}
\end{algorithm}

During the offline training of the CENN, we generate the dataset of $\boldsymbol{c}_{u}$ and $\boldsymbol{g}_{u}$ based on the simulated mmWave channel environment. With the beamspace channel amplitude in \eqref{gu} and the correlation of the received signals and the measurement matrix in \eqref{c}, the training data of $\boldsymbol{c}_{u}$ and $\boldsymbol{g}_{u}$ can be obtained. In fact, the process to obtain $\boldsymbol{c}_{u}$ and $\boldsymbol{g}_{u}$  involves the following four steps: \textbf{i)} we randomly generate a channel vector based on the mmWave channel model in \eqref{hu};  \textbf{ii)} we obtain $\boldsymbol{g}_{u}$ based on \eqref{gu}; \textbf{iii)} we compute the received signal vector $\boldsymbol{r}_{u}$ based on \eqref{ru}; \textbf{iv)} we obtain the correlation vector $\boldsymbol{c}_u$ based on \eqref{c}. We divide the data set into the training set and the validation set randomly, where the size of the training set is nine times the size of the validation set. The output of the CENN is $\hat{\boldsymbol{g}}_{u}$.

The training of the CENN aims to minimize the difference between $\hat{\boldsymbol{g}}_{u}$ and $\boldsymbol{g}_{u}$. The difference, typically named as the loss in machine learning, can be calculated in several
ways. In this work, we calculate the loss by measuring the mean square error as~\cite{ye2018power}
\begin{equation}\label{lossCS}
f_{\rm LossCS}(\boldsymbol{g}_{u},\hat{\boldsymbol{g}}_{u}) = \frac{1}{G} \sum_{n=1}^{G}   \left( \boldsymbol{g}_{u}[n] - \hat{\boldsymbol{g}}_{u}[n]  \right)^2.
\end{equation}

We adopt the adaptive moment estimation (Adam) optimizer to train the CENN by TensorFlow. The CENN is trained
for 1,000 epochs, where 50 mini-batches are utilized in each epoch. The learning rate is set to be a step function, which decreases with training epochs. The learning rate is initialized with the  value of 0.01 and decreases 5-fold every 400 epochs.

During the online deployment of the CENN, we obtain the real measured $\boldsymbol{r}_{u}$ from practical mmWave channel environments. We compute $\boldsymbol{c}_{u}$ based on \eqref{c}, which is then fed to the offline-trained CENN. The prediction of $\boldsymbol{g}_{u}$ by the CENN is $\hat{\boldsymbol{g}}_{u}$.

\subsection{Channel Reconstruction}
Note that the sparsity of $\boldsymbol{h}_{u}^{b}$ can be impaired by channel power leakage caused by the limited beamspace resolution of $\boldsymbol{A}$~\cite{brady2013beam}, which indicates that $\boldsymbol{h}_{u}^{b}$ is not perfectly sparse and  many  entries of $\boldsymbol{h}_{u}^{b}$ have small but nonzero values. Denote the number of dominant entries of  $\boldsymbol{g}_{u}$ by $J$, which is the beamspace channel sparse level. In the
online deployment stage, we sort $\hat{\boldsymbol{g}}_{u}$ in descending order according to the absolute value of $\hat{\boldsymbol{g}}_{u}$. Then we obtain the indices of the first $J$ entries, which are the prediction of the indices of  $J$ dominant entries  in $\boldsymbol{g}_{u}$.

We denote the prediction of these $J$ indices by $\boldsymbol{\Gamma} \in{\mathbb{R}^{J}}$. We further let $\hat{\boldsymbol{h}}_{u}^{b}$ denote an estimate of $\boldsymbol{h}_{u}^{b}$. We initialize $\hat{\boldsymbol{h}}_{u}^{b}$ to be zero. Then the $J$ dominant entries of $\hat{\boldsymbol{h}}_{u}^{b}$ can be computed via the least squares (LS) estimation as
\begin{equation}\label{14}
\hat{\boldsymbol{h}}_{u}^{b}[\boldsymbol{\Gamma}]=(\boldsymbol{\Phi}^H_{\Gamma}\boldsymbol{\Phi}_{\Gamma})^{-1} \boldsymbol{\Phi}^H_{\Gamma} \boldsymbol{r}_{u}
\end{equation}
where $\boldsymbol{\Phi}_{\Gamma}$ consists of $J$ columns of $\boldsymbol{\Phi}$ and the column indices are denoted by $\boldsymbol{\Gamma}$. Then using the result $\boldsymbol{A}^H\boldsymbol{A}=G\boldsymbol{I}_{N_A} / N_A$, we obtain the estimated channel vector for the $u$th user based on (\ref{ru2}) as
\begin{equation}\label{15}
\hat{\boldsymbol{h}}_{u}=\frac{N_A}{G} \boldsymbol{A}^H \hat{\boldsymbol{h}}_{u}^{b}.
\end{equation}

It is shown that the proposed DLCS channel estimation scheme can avoid the greedy search that is commonly adopted by the existing sparse channel estimation schemes based on CS, since the DLCS scheme estimates dominant entries simultaneously instead of sequentially.

\section{DLQP Hybrid Precoder Design}
\begin{figure*}[!t]
\centering
\includegraphics[width=130mm]{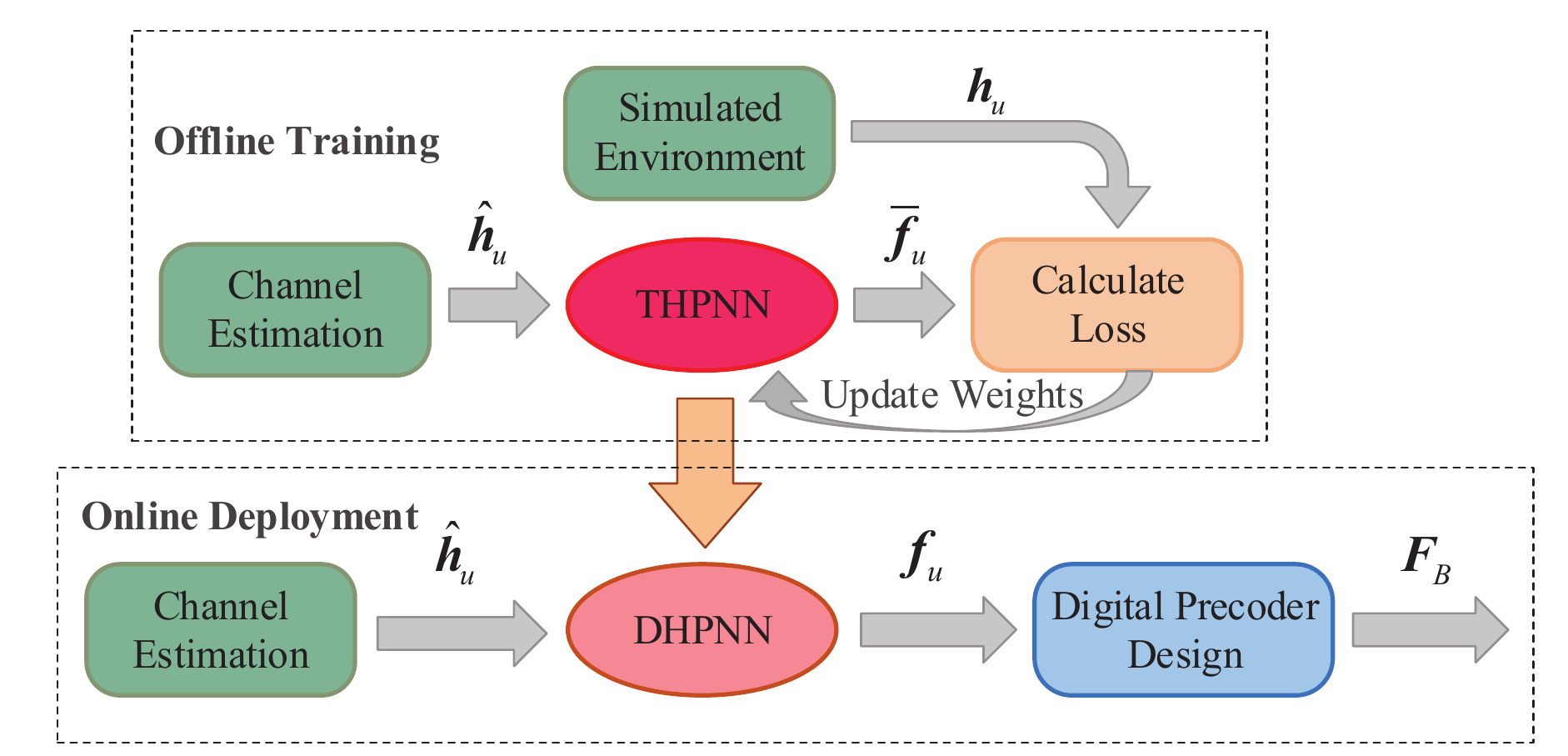}
\caption{Block diagram of the DLQP hybrid precoder design: offline training and online deployment}
\label{FIG9}
\end{figure*}

Hybrid precoding is usually required for downlink data transmission after the channel estimation. In the proposed DLQP hybrid precoder design method, we first design the analog precoder and then the digital precoder. The main idea of the DLQP scheme is  to first train the THPNN using the estimated channel vectors, where the approximate phase quantization is considered.  Then the DHPNN is obtained by replacing the approximate phase quantization in THPNN with ideal phase quantization, where the estimated channel vectors are fed into the DHPNN to obtain the analog precoder vectors. Finally  the analog precoding matrix is obtained by stacking the analog precoding vectors of all users and the digital precoding matrix can be calculated by ZF. The block diagram of the DLQP method is illustrated in Fig.~\ref{FIG9}. The detailed steps of  the DLQP method is summarized in Algorithm~\ref{alg2}.

\subsection{Analog Precoder Design}
\begin{figure*}[!t]
\centering
\includegraphics[width=130mm]{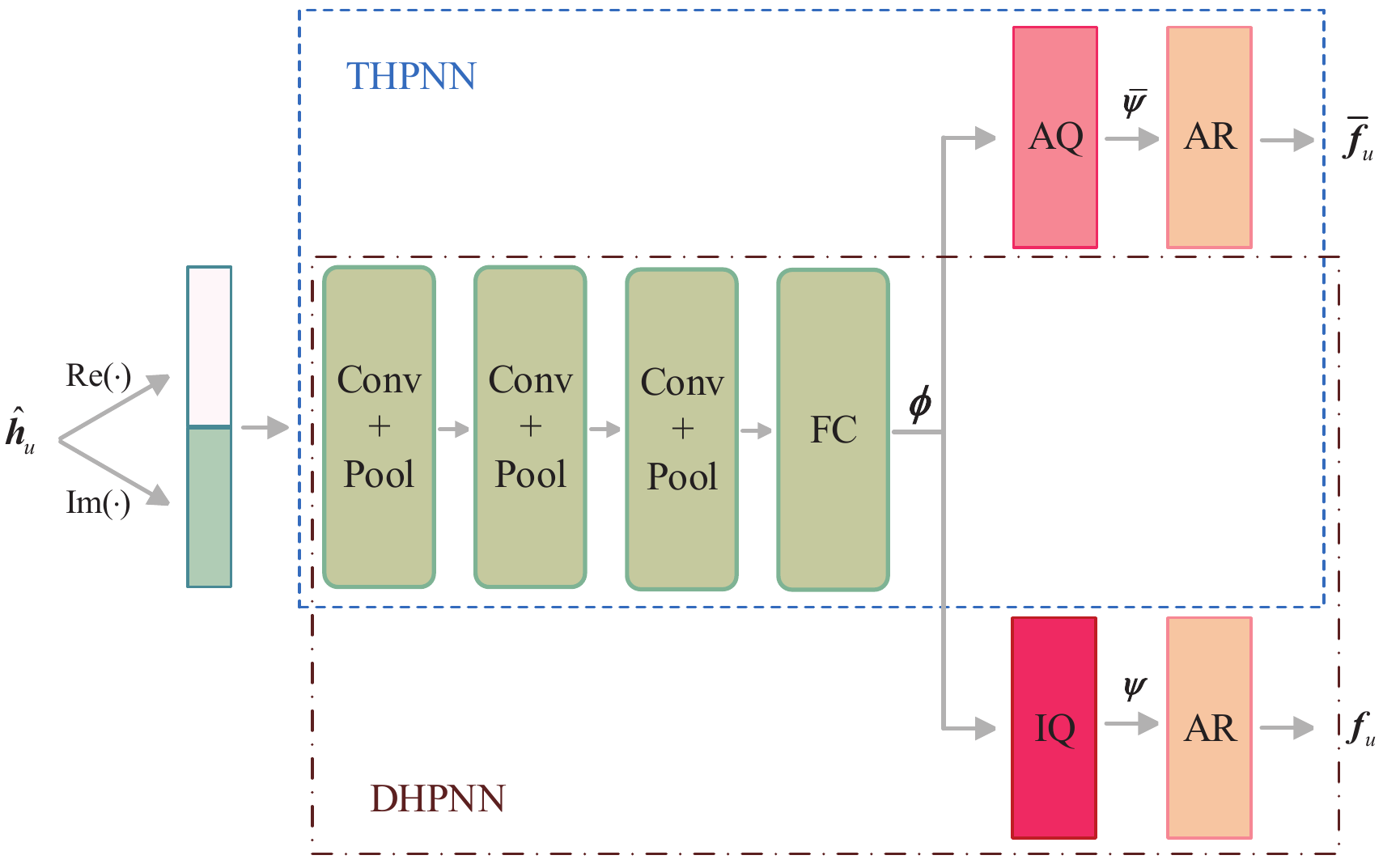}
\caption{Illustration of the THPNN for offline training  and DHPNN for online deployment}
\label{FIG8}
\end{figure*}

Denote the analog precoder vector and approximate analog precoder vector by $\boldsymbol{f}_{u} \triangleq [f_{u,1}, f_{u,2},$ $\ldots, f_{u,N_A}]^T \in{\mathbb{C}^{N_A}}$ and $\bar{\boldsymbol{f}}_{u} \triangleq [\bar{f}_{u,1}, \bar{f}_{u,2}, \ldots, \bar{f}_{u,N_A}]^T \in{\mathbb{C}^{N_A}}$, respectively, for $ u=1,2,\ldots,U$. As shown in Fig.~\ref{FIG8}, $\hat{\boldsymbol{h}}_{u}$ is fed to the THPNN to obtain $\bar{\boldsymbol{f}}_{u}$, while $\hat{\boldsymbol{h}}_{u}$ is fed to the DHPNN to obtain $\boldsymbol{f}_{u}$. Note that the difference between the THPNN and DHPNN is that we use approximate phase quantization in
the THPNN so that the NN can be trained, while we use ideal phase quantization in the DHPNN to meet the practical constraint of limited phase shifter resolution.

We define $B$ as the quantization bit number of the phase shifters used at the BS, where the RF phase is quantized into $Q \triangleq 2^B$ discrete values. Each entry of $\boldsymbol{f}_u$ is randomly drawn from the set $\{ e^{j2\pi n/Q}, n = 1,2,\ldots,Q \}$. The hybrid precoder design schemes based on beamsteering codebooks design the analog precoder vector as the steering vector of the LOS channel path~\cite{zhao2017multi,alk2015limited,sun2019beam}. However, such schemes require that $Q \geq N_A$ to obtain high beamforming gain, which will have unsatisfactory performance when $Q < N_A$. This requirement motivates us to use the NN to design the analog precoder when $Q < N_A$.

\begin{algorithm}[!t]
	\caption{ DLQP Hybrid Precoder Design}
	\label{alg2}
	\begin{algorithmic}[1]
		\STATE \emph{Input:} $\hat{\boldsymbol{h}}_{u}$.
        \vspace{5 pt}
        \STATE \emph{(Analog Precoder Design)}
        \STATE Replace the AQ layer of the offline-trained THPNN by the IQ layer to obtain the DHPNN.
        \STATE Input $\hat{\boldsymbol{h}}_{u}$ to the  DHPNN to get $\boldsymbol{f}_{u}$.
        \STATE Obtain $\boldsymbol{F}_R$ according to (\ref{FR}).
        \vspace{5 pt}
        \STATE \emph{(Digital Precoder Design)}
        \STATE Compute $\hat{\boldsymbol{H}}$ based on (\ref{hatH}).
        \STATE Obtain $\boldsymbol{H}_{\rm{eff}}$ according to (\ref{Heff}).
        \STATE Compute $\boldsymbol{F}_B$ via (\ref{FB}).
        \STATE Normalize each column of $\boldsymbol{F}_B$ via (\ref{fBu}).
        \vspace{5 pt}
        \STATE \emph{Output:} $\boldsymbol{F}_R$, $\boldsymbol{F}_B$.
	\end{algorithmic}
\end{algorithm}

As shown in Fig.~\ref{FIG9}, the hybrid precoder design has two stages: the offline training of the THPNN and online deployment of the DHPNN, where the DHPNN is obtained based on the THPNN by replacing one layer of the THPNN. The THPNN is first trained offline and then the DHPNN is obtained, which is used as the kernel of the hybrid precoder design. The input of the THPNN and DHPNN is $\hat{\boldsymbol{h}}_{u}$. The outputs of the DHPNN and the THPNN are the analog precoder vector $\boldsymbol{f}_{u}$ and approximate analog precoder vector $\bar{\boldsymbol{f}}_{u}$, respectively.

\begin{figure}[!t]
\centering
\includegraphics[width=90mm]{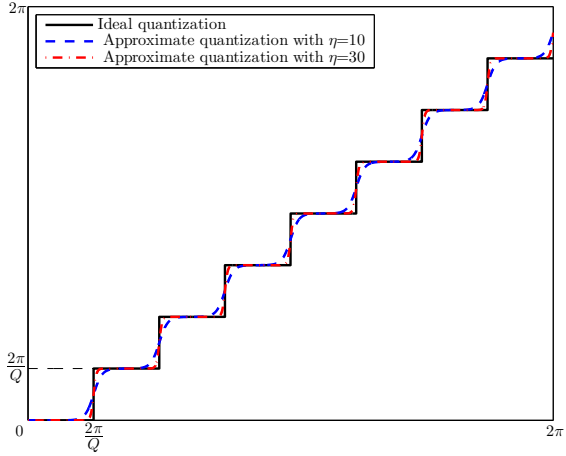}
\caption{Illustration of the ideal quantization and approximate quantization}
\label{FIG10}
\end{figure}

As illustrated in Fig.~\ref{FIG8}, both the adopted THPNN and DHPNN consist of six layers, where five of them are shared. Since the NN can only deal with the real number, the input of the THPNN and DHPNN is a real-valued vector having $2N_A$ entries composed by the imaginary and real  parts of $\hat{\boldsymbol{h}}_{u}$. Each of the first four layers consists of a convolutional (Conv) layer and a pooling (Pool) layer. The kernel size and strides of each Conv layer are set to be five and one, respectively. The number of filters of these three Conv layers are set as 16, 32, and 64, respectively. Both the pool size and strides of each Pool layer are set to be two. The activation function adopted in the first three layers is the ReLU function, while that adopted in the fourth layer is the Sigmoid function, which can be represented as  $f_{\rm Sig}(x)=\frac{1}{1+e^{-x}}$. Since the output of the FC and Pool layers can only be real number, we cannot directly obtain the complex-valued $\boldsymbol{f}_{u}$. Then the output of the fourth layer is the phase of analog precoder vector, which is denoted by
\begin{equation}
\boldsymbol{\phi} \triangleq [\phi_1, \phi_2, \ldots, \phi_{N_A}]^T \in{\mathbb{R}^{N_A}}
\end{equation}
where $\phi_n \in [0,2\pi)$, for $ n=1,2,\ldots,N_A$. Since the RF phase is quantized into $Q$ discrete values, in the DHPNN we use the ideal quantization (IQ) layer to quantize the continuous phase vector $\boldsymbol{\phi}$ into the discrete phase vector. Denote the IQ function and the step function by $\Lambda(\cdot)$ and $\varepsilon (\cdot)$, respectively, where $ \varepsilon (x)=\left\{
\begin{aligned}
 &  0, x<0, \\
&  1, x\geq 0.
\end{aligned}
\right.
$ Then $\Lambda(\cdot)$ can be written as
\begin{equation}
\Lambda(x) \triangleq \frac{2\pi}{Q}\sum_{q=1}^{Q} \varepsilon\left(x - \frac{2\pi q}{Q}\right),~~ x \in [0,2\pi).
\end{equation}
It is shown in Fig.~\ref{FIG10} that $\Lambda(x)$ is not differentiable when $x = 2\pi q / Q$, $q = 1,2,\ldots,Q$, which indicates that standard deep learning training algorithms, such as
stochastic gradient descent (SGD), cannot be directly applied to train the NN. To overcome this problem, we use the approximate  quantization (AQ) layer in the THPNN for offline training instead of the IQ layer. Therefore the DLQP hybrid precoder design method uses two NNs. However, the DLCS channel estimation scheme needs no quantization. Therefore the DLCS channel estimation scheme uses only one NN. Denote the AQ function by $\Gamma(x)$~\cite{rade1998math}, which can be represented as
\begin{equation}
\Gamma(x) \triangleq \frac{\pi}{Q}\sum_{q=1}^{Q} \tanh\left(\eta\left(x - \frac{2\pi q}{Q}\right)\right)+1,~~ x \in [0,2\pi)
\end{equation}
where $\eta$ a constant to represent the degree of approximation. As shown in Fig.~\ref{FIG10}, it is more accurate for $\Gamma(x)$ to approximate $\Lambda(x)$ if we set $\eta$ as a larger number. It is also shown that $\Gamma(x)$ is differentiable for $x \in [0,2\pi)$. Then we use $\Gamma(x)$ to quantize $\boldsymbol{\phi}$ in the THPNN for offline training. Denote the phase vector after quantization by $\bar{\boldsymbol{\psi}} \triangleq [\bar{\psi}_1, \bar{\psi}_2, \ldots, \bar{\psi}_{N_A}]^T \in{\mathbb{R}^{N_A}}$, which can be represented as
\begin{equation}
\bar{\psi}_n = \Gamma(\phi_n), n=1,2,\ldots,N_A.
\end{equation}
Based on the phase vector $\bar{\boldsymbol{\psi}}$, the approximate analog precoder vector $\bar{\boldsymbol{f}}_{u}$ can be obtained in the analog precoder reconstruction (AR) layer. By setting $\bar{\psi}_n$ as the phase of $\bar{f}_{u,n}$, $\bar{f}_{u,n}$ can be represented as
\begin{equation}
\bar{f}_{u,n} = e^{j \bar{\psi}_n}, n=1,2,\ldots,N_A.
\end{equation}

During the offline training of the THPNN, we generate the dataset of $\hat{\boldsymbol{h}}_{u}$ and $\boldsymbol{h}_{u}$ based on the output of the CENN and simulated mmWave channel environment. With the channel vector in \eqref{hu} and the estimated channel vector in \eqref{15}, the training data of $\hat{\boldsymbol{h}}_{u}$ and $\boldsymbol{h}_{u}$ can be obtained. In fact, the process to obtain $\hat{\boldsymbol{h}}_{u}$ and $\boldsymbol{h}_{u}$  involves the following five steps: \textbf{i)} we randomly generate the channel vector $\boldsymbol{h}_{u}$ based on the mmWave channel model in \eqref{hu};  \textbf{ii)} we compute the received signal vector $\boldsymbol{r}_{u}$ based on \eqref{ru}; \textbf{iii)} we obtain the correlation vector $\boldsymbol{c}_u$ based on \eqref{c}; \textbf{iv)} we feed $\boldsymbol{c}_u$ to the offline-trained CENN for the DLCS channel estimation to get $\hat{\boldsymbol{g}}_{u}$; \textbf{v)} we obtain  the estimated channel vector $\hat{\boldsymbol{h}}_{u}$ based on the  channel reconstruction in \eqref{15}.  The output of the THPNN is $\bar{\boldsymbol{f}}_{u}$.

The training of the THPNN aims to maximize the beamforming gain, i.e., the inner product of $\bar{\boldsymbol{f}}_{u}$ and $\boldsymbol{h}_{u}$. Since the THPNN is trained to minimize the loss, we calculate the loss as the opposite number of the inner product, which can be represented as~\cite{wang2018mmwave}
\begin{equation}\label{lossQP}
f_{\rm LossQP}(\bar{\boldsymbol{f}}_{u},\boldsymbol{h}_{u}) = -\left| \bar{\boldsymbol{f}}_{u}^T \boldsymbol{h}_{u}  \right|.
\end{equation}

The training of the CENN aims to minimize the difference between $\hat{\boldsymbol{g}}_{u}$ and $\boldsymbol{g}_{u}$, while the training of the THPNN aims to maximize the beamforming gain. Therefore the loss function in \eqref{lossCS} is different from that in \eqref{lossQP}. Note that the output of the NN is the  analog precoder
vector $\bar{\boldsymbol{f}}_u$, while we need to calculate the analog precoding matrix $\boldsymbol{F}_R$ so that the spectral efficiency can be obtained. However, the computational process from $\bar{\boldsymbol{f}}_u$ to $\boldsymbol{F}_R$ is not differentiable, which cannot be applied to the NN. Therefore we do not set the spectral efficiency as the loss. We adopt the adaptive moment estimation (Adam) optimizer to train the THPNN by TensorFlow. The THPNN is trained
for 6,000 epochs, where 200 mini-batches are utilized in each epoch. The learning rate is set to be a step function. The learning rate is initialized with the  value of 0.01 and decreases 2-fold every 2000 epochs.

During the online deployment stage, we obtain the DHPNN based on the offline-trained THPNN by replacing the AQ layer in the THPNN with IQ layer. To obtain the input of the THPNN $\hat{\boldsymbol{h}}_{u}$, we obtain the real measured $\boldsymbol{r}_{u}$ from practical mmWave channel environments. We compute $\boldsymbol{c}_{u}$ based on \eqref{c}, which is then fed to the offline-trained CENN for the DLCS channel estimation to get $\hat{\boldsymbol{g}}_{u}$.  We obtain  the estimated channel vector $\hat{\boldsymbol{h}}_{u}$ based on the  channel reconstruction in \eqref{15}. We then feed $\hat{\boldsymbol{h}}_{u}$ to the DHPNN for the DLQP hybrid precoder design to get $\boldsymbol{f}_u$. Note that different from the offline training of the THPNN, we use the IQ function $\Lambda(x)$ to quantize $\boldsymbol{\phi}$ in the DHPNN, which ensures that each entry of $\boldsymbol{f}_u$ is drawn from the set $\{ e^{j2\pi n/Q}, n = 1,2,\ldots,Q \}$. Denote the phase vector after quantization by $\boldsymbol{\psi} \triangleq [\psi_1, \psi_2, \ldots, \psi_{N_A}]^T \in{\mathbb{R}^{N_A}}$, which can be represented as
\begin{equation}
\psi_n = \Lambda(\phi_n), n=1,2,\ldots,N_A.
\end{equation}
Based on the phase vector $\boldsymbol{\psi}$, the analog precoder vector $\boldsymbol{f}_{u} $ can be obtained. By setting $\psi_n$ as the phase of $f_{u,n}$, $f_{u,n}$ can be represented as
\begin{equation}
f_{u,n} = e^{j \psi_n}, n=1,2,\ldots,N_A.
\end{equation}

After obtaining $\boldsymbol{f}_{u} $ in the online deployment stage, the analog precoding matrix $\boldsymbol{F}_R$ can be represented as
\begin{equation}\label{FR}
\boldsymbol{F}_R = [\boldsymbol{f}_{1},\boldsymbol{f}_{2},\ldots,\boldsymbol{f}_{U}].
\end{equation}

It is shown in \eqref{lossQP} that the analog precoder is designed to maximize the beamforming gain, where the quantization of the RF phase is considered. Note that although we use the AQ layer for offline training, we adopt the IQ layer in the online deployment stage, which guarantees the consistency of our adopted NN and the practical hardware constraint of limited phase shifter resolution.

\subsection{Digital Precoder Design}
We denote the estimated channel matrix for the BS and all users by
\begin{equation}\label{hatH}
\hat{\boldsymbol{H}} \triangleq \left[\hat{\boldsymbol{h}}_{1},...,\hat{\boldsymbol{h}}_{U}\right]^{T} \in{\mathbb{C}^{U\times{N_A}}}.
\end{equation}
We further denote the effective channel matrix  by
\begin{equation}\label{Heff}
\boldsymbol{H}_{\rm{eff}} \triangleq \hat{\boldsymbol{H}} \boldsymbol{F}_R.
\end{equation}
Analog precoding aims to form directional beams using phase shifter network, while digital precoding is designed to mitigate interference of multiple data
streams after analog precoding. Then the ZF digital precoding matrix can be represented by
\begin{equation}\label{FB}
\boldsymbol{F}_B = \boldsymbol{H}_{\rm{eff}}^H (\boldsymbol{H}_{\rm{eff}} \boldsymbol{H}_{\rm{eff}}^H)^{-1}.
\end{equation}
To satisfy the total power constraint, each column of the designed digital precoder, denoted by $\boldsymbol{f}_{B,u}$, should be normalized, i.e.,
\begin{equation}\label{fBu}
\boldsymbol{f}_{B,u} = \boldsymbol{f}_{B,u} / \left\| \boldsymbol{F}_R \boldsymbol{f}_{B,u} \right\|_2
\end{equation}
such that $\left\| \boldsymbol{F}_R \boldsymbol{f}_{B,u} \right\| _2 ^2= 1$, $u=1,2,\ldots,U$.

It is shown that the proposed DLQP hybrid precoder design method can obtain the analog precoder considering the  quantized  phase constraint, which is of great  value in practical mmWave systems.

\section{Simulation Results}
In the following we will present the performance evaluation for the proposed DLCS channel estimation scheme and the proposed DLQP hybrid precoder design method. Considering a multi-user mmWave massive MIMO communication system, the BS equipped with $N_R=4$ RF chains and $N_A=64$ antennas serves $U=3$  users with single antenna.  We set $G=128$ according to~\cite{javier2018frequency}, and we set the number of multiple paths in mmWave channel as $L_{u}=2$, where $g_{u,1}\thicksim\mathcal{CN}(0,1)$ and $g_{u,2}\thicksim\mathcal{CN}(0,0.5)$~\cite{javier2018frequency,chen2020two}. For the uplink pilot transmission, we set $\boldsymbol{F}_B^k = \boldsymbol{I}_{N_R}$. Therefore the hybrid precoding matrix is equal to the analog precoding matrix and is also a random matrix. The quantization bit number of the phase shifters used at the BS is $B=4$, leading to $Q=16$~\cite{javier2018frequency}. We set $\eta=100$. Since $\boldsymbol{h}_{u}^b$ is not ideally sparse due to the power leakage,  the beamspace channel sparse level should be larger than $L_{u}$, i.e., $J>2$. We set $J=6,7$ in performance simulating. Note that the CENN is trained to predict the beamspace channel amplitude, where the training process of the CENN is independent of $J$.   The proposed DLCS  channel estimation scheme is compared with the existing OMP~\cite{venugopal2017time} and DGMP~\cite{dai2016estimation} channel estimation schemes, while the proposed DLQP hybrid precoder design method is compared with the existing QALS~\cite{zhao2017multi} hybrid precoder design method. We also compare the DLQP method with the Exhaustion hybrid precoder design method, i.e., we generate the analog precoding matrix $\boldsymbol{F}_R$ for $30,000$ times, where each entry of $\boldsymbol{F}_R$ is randomly drawn from the set $\{ e^{j2\pi n/Q}, n = 1,2,\ldots,Q \}$, and then the digital precoder is designed according to Algorithm~\ref{alg2}. We select the hybrid precoder with the largest spectral efficiency as the output of the Exhaustion hybrid precoder design method.

\begin{figure}[!t]
\centering
\includegraphics[width=90mm]{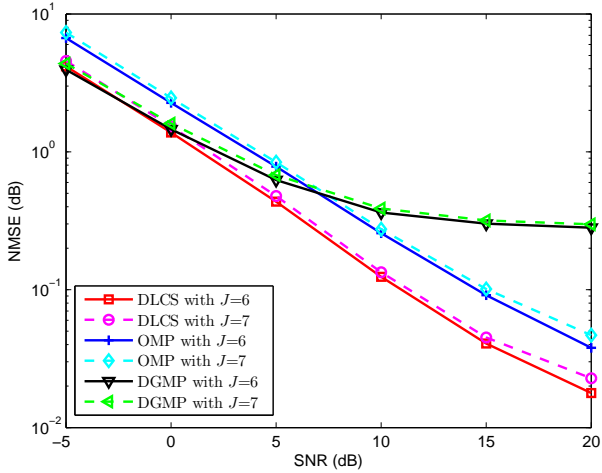}
\caption{Comparisons of channel estimation performance in terms of SNR for different schemes}
\label{Fig4}
\end{figure}

\begin{figure}[!t]
\centering
\includegraphics[width=90mm]{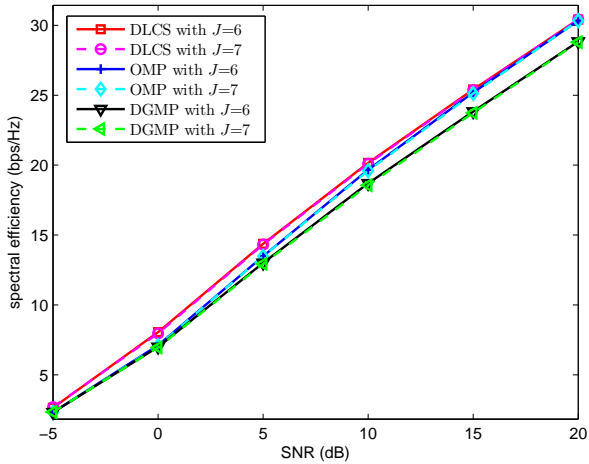}
\caption{Comparisons of spectral efficiency in terms of SNR for different channel estimation schemes}
\label{Fig5}
\end{figure}

\begin{figure}[!t]
\centering
\includegraphics[width=90mm]{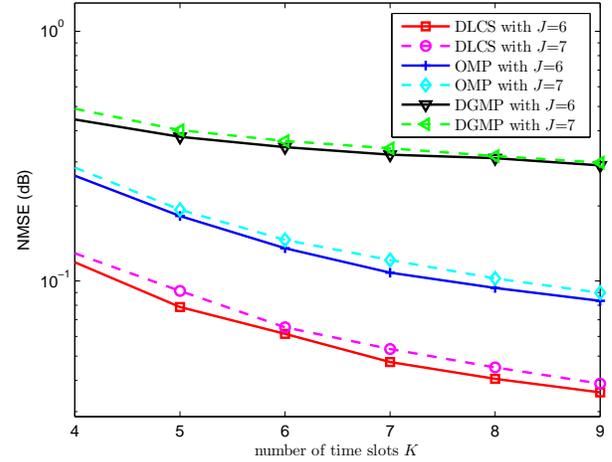}
\caption{Comparisons of channel estimation performance in terms of the number of   time slots for channel training for different schemes}
\label{FIG6}
\end{figure}

\begin{figure}[!t]
\centering
\includegraphics[width=88mm]{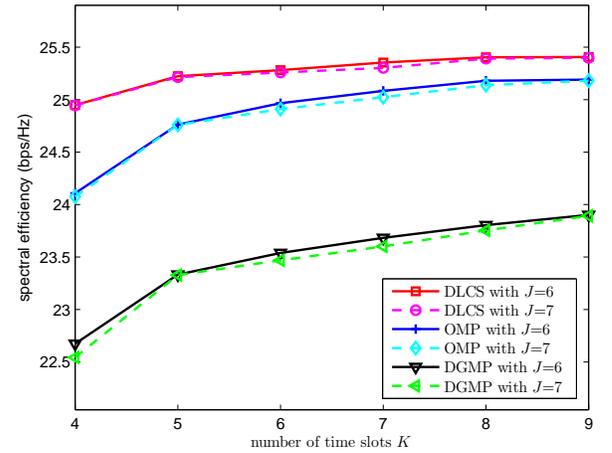}
\caption{Comparisons of spectral efficiency in terms of the number of   time slots for channel training for different channel estimation schemes}
\label{FIG7}
\end{figure}

We first evaluate the performance of the proposed DLCS channel estimation scheme from Fig.~\ref{Fig4} to Fig.~\ref{FIG7}. As shown in Fig.~\ref{Fig4}, the channel estimation performance for the proposed DLCS scheme together with the existing schemes is compared in terms of SNR. The channel estimation performance is measured by the normalized mean-squared error (NMSE), which is defined by
\begin{equation}
  \textrm{NMSE} \triangleq \frac{ \sum_{u=1}^{U} \| \hat{\boldsymbol{h}}_{u} - \boldsymbol{h}_{u} \|_2^2 }{ \sum_{u=1}^{U}  \| \boldsymbol{h}_{u} \|_2^2 }.
\end{equation}

We use $K=8$ time slots to transmit pilots for uplink channel estimation.  To make a fair comparison, we fix the pilot training  time slots  to be  eight for the OMP  and DGMP  schemes. It is shown that  the DLCS  scheme has better channel estimation performance than existing schemes. When SNR = 10 dB, the DLCS  scheme with $J=6$ has 51.7\% and 65.8\% performance improvements over the OMP  and DGMP  schemes, respectively, while the DLCS  scheme with $J=7$ has 51.3\% and 65.5\% performance improvements over the OMP  and DGMP  schemes, respectively. We explain the reason for the performance improvements as follows. The OMP scheme estimates the beamspace channel dominant entries sequentially, which cannot guarantee global optimality. The DGMP  scheme only estimates the LOS path, while the proposed DLCS scheme can simultaneously estimate all the dominant beamspace channel entries.

As shown in Fig.~\ref{Fig5}, we compare the spectral efficiency for the proposed DLCS scheme with the existing schemes in terms of SNR. Based on the estimated channel, there are various methods to design the hybrid precoding for mmWave downlink transmission. Similar to ~\cite{javier2018frequency}, in this work we wish to compare the upper bound of the downlink  spectral efficiency, which can be simply measured by the fully-digital precoding.  The ZF precoding matrix can be represented by $\boldsymbol{F}^{\rm{dl}} \triangleq (\hat{\boldsymbol{H}}^* \hat{\boldsymbol{H}}^T)^{-1} \hat{\boldsymbol{H}}^*$. To meet the total power budget, the $u$th row of $\boldsymbol{F}^{\rm{dl}}$, denoted by $\boldsymbol{f}_u^{\rm{dl}}$, should be normalized, i.e., $\boldsymbol{f}_u^{\rm{dl}} \leftarrow \boldsymbol{f}_u^{\rm{dl}} / \| \boldsymbol{f}_u^{\rm{dl}} \|_2$ such that $\| \boldsymbol{f}_u^{\rm{dl}} \|_2=1$ for $u=1,2,\ldots,U$.  Then the spectral efficiency is given by~\cite{javier2018frequency}
\begin{equation}
    R \triangleq \sum_{u=1}^{U} \log_2\left( 1 + \frac{ \frac{1}{U} \left| \boldsymbol{f}_u^{\rm{dl}} \boldsymbol{h}_{u} \right|^2 }{ \frac{1}{U} \sum_{i\neq u} \left| \boldsymbol{f}_i^{\rm{dl}} \boldsymbol{h}_{u} \right|^2 + \sigma^{2} } \right).
\end{equation}

It is seen from Fig.~\ref{Fig5} that the DLCS scheme has better channel estimation performance than existing schemes. When SNR = 10 dB, the DLCS  scheme with $J=6$ has 2.5\% and 7.8\% performance improvements over the OMP  and  DGMP  schemes, respectively, while the DLCS  scheme with $J=7$ has 2.6\% and 8.3\% performance improvements over the OMP  and  DGMP  schemes, respectively. The reason for the smaller spectral efficiency gap between different schemes than the NMSE gap is that the NMSE performance is much more sensitive to the success rate of the sparse recovery, while the spectral efficiency performance is determined by the beamforming gain and is less sensitive to the success rate of the sparse recovery.

In Fig.~\ref{FIG6}, the channel estimation performance for the DLCS, OMP, and DGMP  schemes is compared in terms of the number of  time slots for channel training.   We use the same number of pilot training  time slots for the DLCS, OMP, and DGMP  schemes. SNR is fixed as 15 dB. From Fig.~\ref{FIG6}, it is shown that the DLCS scheme has the best channel estimation performance. When fixing the number of pilot training  time slots to be $K=7$, the DLCS  scheme with $J=6$ has 56.1\% and 85.2\% performance improvements over the OMP  and  DGMP  schemes, respectively, while the DLCS  scheme with $J=7$ has 56.0\% and 84.3\% performance improvements over the OMP  and  DGMP  schemes, respectively.

As shown in Fig.~\ref{FIG7}, we compare the spectral efficiency for different schemes in terms of the number of time slots for channel training. The system parameters for performance simulation are set to be the same as those for Fig.~\ref{FIG6}. It is shown that the DLCS  scheme can have better channel estimation performance than the OMP  and  DGMP  schemes. When the number of channel training  time slots   is more than eight, the spectral efficiency of the DLCS  scheme remains constant, indicating that $K=8$ is sufficient to obtain the full channel state information.

\begin{figure}[!t]
\centering
\includegraphics[width=90mm]{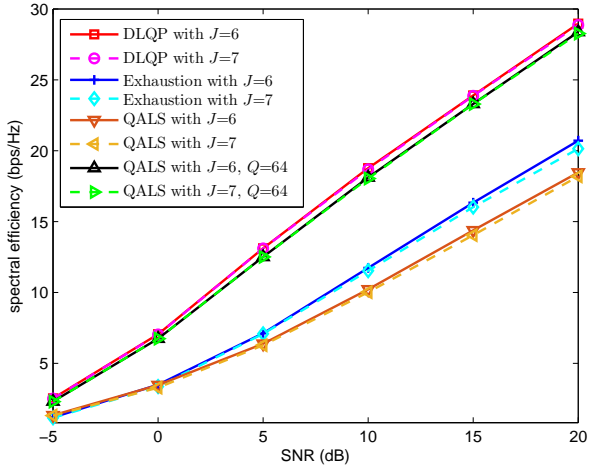}
\caption{Comparisons of spectral efficiency in terms of SNR for different hybrid precoder design methods}
\label{Fig11}
\end{figure}

\begin{figure}[!t]
\centering
\includegraphics[width=90mm]{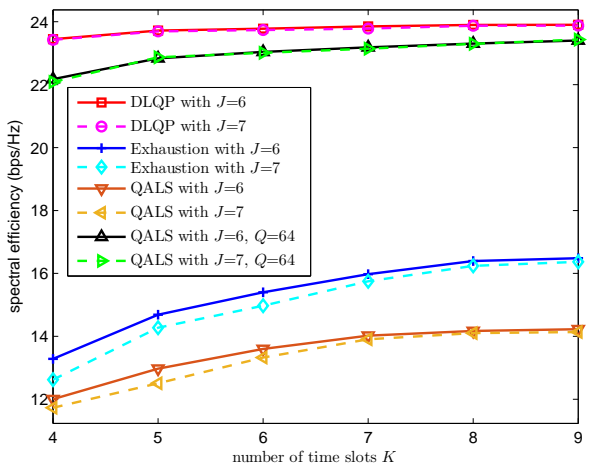}
\caption{Comparisons of spectral efficiency in terms of the number of time slots for channel training for different hybrid precoder design methods}
\label{Fig12}
\end{figure}

In the following, we evaluate the performance of the proposed DLQP hybrid precoder design method in Fig.~\ref{Fig11} and Fig.~\ref{Fig12}. Fig.~\ref{Fig11} compares of the spectral efficiency for the proposed DLQP hybrid precoder design method together with the existing methods in terms of SNR. Since the QALS method needs high phase shifter resolution to obtain  analog beamforming vectors aligning with
the dominant channel paths, we also simulate the QALS method with $Q=64$. It is seen from Fig.~\ref{Fig11} that the DLQP method has better spectral efficiency performance than existing methods. When SNR = 10 dB, the DLQP  method with $J=6$ has 59.9\%, 83.6\% and 3.5\% performance improvements over the Exhaustion, QALS with $Q=16$  and  QALS with $Q=64$  methods, respectively, while the DLQP  method with $J=7$ has 62.0\%, 86.5\% and 3.6\% performance improvements over the Exhaustion, QALS with $Q=16$  and  QALS with $Q=64$  methods, respectively. We explain the reason for the performance gap as follows. In the Exhaustion method, although we generate the analog precoding matrix $\boldsymbol{F}_R$ for $30,000$ times, the number of the total possible $\boldsymbol{F}_R$ should be $Q^{N_A U}=1.55 \times 10^{231}$, which is far more than the acceptable  computational complexity. In the QALS method, the AoA of the LOS channel path cannot be aligned with well with  the small number of available steering vectors of quantized angles.

As shown in Fig.~\ref{Fig12}, we compare the spectral efficiency for different hybrid precoding methods in terms of the number of time slots for channel training. The system parameters for performance simulation are set to be the same as those for Fig.~\ref{FIG6}. It is shown that the DLQP  method can have better spectral efficiency performance than the Exhaustion and QALS  methods. When fixing the number of pilot training  time slots to be $K=7$, the DLQP  method with $J=6$ has 49.3\%, 70.1\% and 2.9\% performance improvements over the Exhaustion, QALS with $Q=16$  and  QALS with $Q=64$  methods, respectively, while the DLQP  method with $J=7$ has 51.0\%, 71.0\% and 2.7\% performance improvements over the Exhaustion, QALS with $Q=16$  and  QALS with $Q=64$  methods, respectively.

\section{Conclusions}
We proposed a DLCS  channel estimation scheme and a DLQP hybrid precoder design method for the multi-user mmWave massive MIMO communication systems. The proposed DLCS scheme and DLQP method were compared with the existing  works in the aspect of NMSE and spectral efficiency. Simulation results showed that the proposed DLCS scheme has better channel estimation performance than existing schemes and the proposed DLQP method has high spectral efficiency with low resolution of phase shifters. As a future work, it is worth developing the channel estimation and hybrid precoding design for wideband multi-user mmWave massive MIMO transmission adopting deep learning.

\bibliographystyle{IEEEtran}
\bibliography{IEEEabrv,IEEEexample}

\end{document}